\pdfoutput=1
\documentclass[a4paper,fleqn,usenatbib]{mnras}
\usepackage[T1]{fontenc}
\usepackage{ae,aecompl}
\usepackage{graphicx}	% Including figure files
\usepackage{amsmath}	% Advanced maths commands
\usepackage{amssymb}	% Extra maths symbols
\usepackage{hyperref}

\title[Efficiency of Geometric Samplers]{The Efficiency of Geometric Samplers for Exoplanet Transit Timing Variation Models}
\author[Tuchow et al.]{Noah W. Tuchow$^{1,2,3}$\thanks{Email: nxt5109@psu.edu},
Eric B. Ford$^{1,2,3,4}$, Theodore Papamarkou$^{5}$  and Alexey Lindo$^{5}$
\\
$^{1}$Department of Astronomy \& Astrophysics, Pennsylvania State University, University Park, PA 16802, USA \\
$^{2}$Center for Exoplanets and Habitable Worlds, Pennsylvania State University, University Park, PA 16802, USA\\
$^{3}$Center for Astrostatistics, Pennsylvania State University, University Park, PA 16802, USA \\
$^{4}$Institute for CyberScience, Pennsylvania State University, University Park, PA 16802, USA\\
$^{5}$School of Mathematics and Statistics, University of Glasgow, University Place, Glasgow G12 8QQ, UK
}
\date{Accepted XXX. Received YYY; in original form ZZZ}

\pubyear{2018}

\begin{document}
\label{firstpage}
\pagerange{\pageref{firstpage}--\pageref{lastpage}}
\maketitle

\begin{abstract}
Transit timing variations (TTVs) are a valuable tool to determine the masses and orbits of transiting planets in multi-planet systems. TTVs can be readily modeled given knowledge of the interacting planets' orbital configurations and planet-star mass ratios, but such models are highly nonlinear and difficult to invert.
Markov chain Monte Carlo (MCMC) methods are often used to explore the posterior distribution for model parameters, but, due to the high correlations between parameters, nonlinearity, and potential multi-modality in the posterior, many samplers perform very inefficiently. Therefore, we assess the performance of several MCMC samplers that use varying degrees of geometric information about the target distribution. We generate synthetic datasets from multiple models, including the TTVFaster model and a simple sinusoidal model, and test the efficiencies of various MCMC samplers. We find that sampling efficiency can be greatly improved for all models by sampling from a parameter space transformed using an estimate of the covariance and means of the target distribution. No one sampler performs the best for all datasets. For datasets with near Gaussian posteriors, the Hamiltonian Monte Carlo sampler obtains the highest efficiencies when the step size and number of steps are properly tuned.  Two samplers --- Differential Evolution Monte Carlo and Geometric adaptive Monte Carlo, have consistently efficient performance for each dataset.  Based on differences in effective sample sizes per time, we show that the right choice of sampler can improve sampling efficiencies by several orders of magnitude.
\end{abstract}
%citations in abstract?
%add celestial mechanics, different tags

\section{Background}
\subsection{Astronomical Background}

One of the main objectives in exoplanet science is to obtain a relationship between planetary masses and radii to determine the average densities of planets and constrain their probable compositions. In order to obtain a mass-radius relation, one would typically need both radial velocity and transit observations for each planet.
In the radial velocity planet detection method, host star spectra are carefully monitored to detect the Doppler shifts caused by the motion of the star-planet system around their barycenter. This detection method determines the minimum mass of a planet required to cause the observed reflexive motion in its star, and allows for characterization of its orbit. Planets found via the transit technique are found via the dimming of their host stars as the planet passes in front of them. This gives us a measurement of the radius of the planet and its orbital period.

However, it is often the case that both techniques cannot be applied to the same system. Many radial velocity systems do not have the precise alignment needed to observe planets passing in front of their star. Similarly, many transiting systems are not suitable for radial velocity follow-up observations due to various causes, such as the faintness or noisiness of their host stars, making it challenging to obtain precise velocity measurements from their spectra.
%making it impractical  for precise spectra to be taken.

Fortunately, there is another method for measuring exoplanet masses and radii for transiting multi-planet systems. The transit timing variation (TTV) method measures the discrepancies between observed times of transits and those projected from a Keplerian model \citep{agol2017Review,holman2010}.  Using the gravitational interactions between transiting planets, one can obtain their masses and eccentricities. Usually, these interactions are small enough that they are only observable with current instruments when the planets are near orbital resonances or very closely spaced. With the vast amount of data obtained by the Kepler mission, roughly 260 systems displaying significant long-term TTVs have been found \citep{lissauer2011,holczer2016}.
However, it is often difficult to determine planetary properties from observed TTVs \citep{jontofhutter2016}. Dynamical models use planetary parameters to compute predicted transit times, but it is nontrivial to invert the problem to obtain planetary parameters from transit times.

Ideally, one would like to use N-body simulations to model the dynamical interactions between planets in TTV systems. For instance, the TTVFast model uses a symplectic N-body integrator to calculate TTVs from a given set of initial conditions \citep*{deck2014}.
Because such N-body models are often computationally intensive and time consuming, an alternative approximate model, TTVFaster, was developed for systems near first order orbital resonances \citep{agol2016}. This semi-analytic model approximates TTVs using a series expansion  and greatly reduces computation times.

\subsection{Statistical Background}
\label{sec:statbackground}

One of the main challenges in characterizing exoplanets with TTVs is that it is difficult to invert TTV models to obtain planetary properties.
Previous studies to determine planetary masses and orbits from TTV data have been very time consuming (e.g. \citealt{carter2012}). In some cases, degenerate solutions for planetary masses and eccentricities were found to yield TTV predictions of similar statistical quality \citep{jontofhutter2016}.
%rephrase?
The posterior distributions for model parameters may also display multimodality, making it even more challenging to explore the parameter space.

 Since even the simplest TTV models require exploring a 10+ dimensional parameter space, characterizing the model parameter space is a nontrivial matter with traditional Markov Chain Monte Carlo (MCMC) methods. The most basic form of MCMC we are concerned with is the Random Walk Metropolis-Hastings algorithm (e.g. \citealt{ford2005}). This method samples from a target distribution by proposing new states of parameter values drawn from a proposal distribution that is centered on the current state.  After each proposal, the prior and likelihood of the proposed state are calculated to determine whether to accept the proposal. The method used for proposing steps in the MCMC chain depends on the specific sampler used. Proposing steps with increased geometric knowledge of the target distribution will often allow one to sample much more efficiently \citep{GirolamiCalderhead}.
 %better transition?
For this reason, several samplers, such as the Metropolis-adjusted Langevin algorithm (MALA), choose a step based on the known gradient of the target distribution \citep{RobertsRosenthal1998}. Similarly, Hamiltonian Monte Carlo (HMC) proposes a state by leapfrog integration of a Hamiltonian dynamical system based on the current position of the MCMC chain and the target gradient \citep{duane1987}. HMC uses two tuning parameters, namely the integration step size and the number of leapfrog sub-steps per proposal.

 There are also samplers designed to use an ensemble of MCMC chains to propose a new state.  Samplers such as the Differential Evolution MCMC (DEMCMC) and the Affine-Invariant ensemble MCMC (AIMCMC) samplers do not require the gradient of the target distribution, but instead approximate the target distribution's shape using the difference in parameter values of random walkers in the ensemble \citep{terbraak2006, goodman2010}. Both of these ensemble samplers have been frequently used in previous studies of exoplanet TTVs (e.g. \citealt{carter2012,jontofhutter2016}), and AIMCMC is common choice of sampler, due in part to its easy to use implementation in the popular \texttt{emcee} python package \citep{emcee}.

Recently, samplers have been designed to use not only gradients, but also the Hessians of the target distribution. The simplified manifold Metropolis-adjusted Langevin algorithm (SMMALA) uses the Hessian of the target distribution to make efficient proposals for every step \citep{GirolamiCalderhead}.
%rephrase, smarter proposals?
 This method utilizes a large amount of geometric information, but computing the Hessian can be time consuming. For this reason, the Geometric adaptive Monte Carlo (GAMC) sampler was developed. GAMC uses the Hessian frequently at first, but gradually reduces the frequency of Hessian evaluations via an exponential schedule \citep*{papamarkou2018}. GAMC is summarized briefly in Appendix \ref{appendix:GAMC}.

\section{Methods}

We would like to develop a more streamlined and computationally efficient means of using MCMC methods to characterize exoplanets based on TTV observations.
For this study, our goal was to compare the computational efficiency of multiple MCMC samplers when applied to characterize the posterior distributions for TTV models. We consider two main classes of models for exoplanet transit times: a simple sinusoidal model (SSM) and the semi-analytical TTVFaster models.

\subsection{Models}

For both TTV models we choose a log-likelihood in the form:
\begin{equation}
\log{\mathcal{L}_m (p|\mathbf{T})}= -\frac{1}{2} \chi_m^2 -\sum_{i \in N_m} \frac{1}{2} \log{(2\pi \sigma_{m,i})}
\end{equation}
where $\chi_m^2 = \sum_{i \in N_m} \left( \frac{\tau_{m,i}-T_{m,i}}{\sigma_{m,i}}\right) ^2 $ with m= 1, 2 denoting the inner and outer planets respectively.
Here $\tau_{m,i}$ are the transit times calculated from the TTV model with parameters $p$, $T_{m,i}$ are the measured transit times, and $\sigma_{m,i}$ are the measurement uncertainties. The set of numbered transits observed for planet m is denoted as $N_m$. The total likelihood is obtained by summing the log-likelihoods, $\log{\mathcal{L}_m}$, for both planets.

In order to use more sophisticated samplers that use geometric information about the target distribution, we need to compute the first and second order partial derivatives of the log-likelihood. For both models, we use forward automatic differentiation with the  \texttt{ForwardDiff.jl} package\footnote{\url{https://github.com/JuliaDiff/ForwardDiff.jl.git}} in Julia to calculate first and second order derivatives \citep*{forwarddiff}.
The SSM model's derivatives can be easily computed analytically or with low memory usage via forward automatic differentiation.
However, for the TTVFaster model, we found that derivatives of the log-likelihood obtained purely via forward automatic differentiation were too slow and computationally inefficient, in some cases causing memory overflow errors. To resolve this problem, we instead used analytic chain rule differentiation to obtain gradients in the form
 \begin{equation}
\frac{\partial}{\partial p_i}  \log{\mathcal{L}_{tot}} =\sum_{m=1,2} \sum_{k \in N_m} \frac{-(\tau_{m,k} -T_{m,k})}{\sigma_{m,k} ^2} \frac{\partial \tau_{m,k}}{\partial p_i}
 \end{equation}
 where $\frac{\partial \tau_{m,k}}{\partial p_i}$ is the Jacobian of the TTVFaster output transit times for planet m with respect to model parameters, evaluated via forward auto-differentiation. Similarly Hessians are obtained via
 \begin{equation}
 \begin{aligned}
 \frac{\partial^2  \log{\mathcal{L}_{tot}}}{\partial p_i  \partial p_j}  = &\sum_{m=1,2}\sum_{k \in N_m} -\Big( \frac{\partial \tau_{m,k}}{\partial p_i} \frac{\partial \tau_{m,k}}{\partial p_j} \\
 & + (\tau_{m,k}-T_{m,k}) \frac{\partial^2 \tau_{m,k}}{\partial p_i \partial p_j}   \Big) / \sigma_{m,k}^2
 \end{aligned}
 \end{equation}
% Note that $k$ here is a dummy index referring to a specific transit number and does not refer to eccentricity vector component $k=e \cos \omega$.

\subsubsection{Simple Sinusoidal Model}

As a stepping stone towards sampling from complicated TTV models, we first aim to understand how to sample efficiently for a simpler, but closely related model. The TTVFaster model calculates transit times from a series expansion around an orbital resonance, so it can ultimately be thought of as the sum of multiple sinusoids \citep{agol2016}.
%clarify?
Earlier models for exoplanet TTVs, such as that of \citet*{lithwickwu2012}, modeled transit times analytically using a sinusoidal waveform.  This motivated us to try a simple sinusoidal model (SSM), modeling transit times for each planet as a linear ephemeris plus a sinusoid at the fundamental frequency and its first harmonic term.
\begin{equation}
\label{SSMeq}
\begin{aligned}
\tau_m (N_m, p) &= t_{lin,m} + \\
                              & A_m \sin{(f_{TTV} t_{lin,m})} +B_m \cos{(f_{TTV} t_{lin,m})} +\\   &C_m \sin{(2 f_{TTV} t_{lin,m})}+ D_m \cos{(2 f_{TTV} t_{lin,m})}
\end{aligned}
\end{equation}
For two planets, this model has 12 parameters $p= \{t_{i,1}, P_1,  A_1, B_1, C_1, D_1, t_{i,2}, P_2, A_2, B_2, C_2, D_2\}$, two more than the TTVFaster model.
In equation \ref{SSMeq}, $t_{lin,m}$ is the linear ephemerus for planet m defined as $t_{lin,m}(N_m,p)=t_{i,m} + (N_m - 1) P_m$, where $P_m$ and $t_{i,m}$ are model parameters and recall $N_m$ is the set of observed transit numbers.
%add footnote about about how t_{i,m} is time of first transit and corresponding to N_m =1
 Here we define $f_{TTV}$  as  $f_{TTV}=2\pi / P_{TTV}$, and $P_{TTV}$ is the superperiod of TTV signals for a pair of planets near a $\beta:\alpha$ resonance given by $P_{TTV} = (\beta/P_2 - \alpha/P_1 )^{-1}$.
%specify first order \beta=\alpha+1

%modified: gave amount of noise
\begin{figure}
\includegraphics[width=\columnwidth]{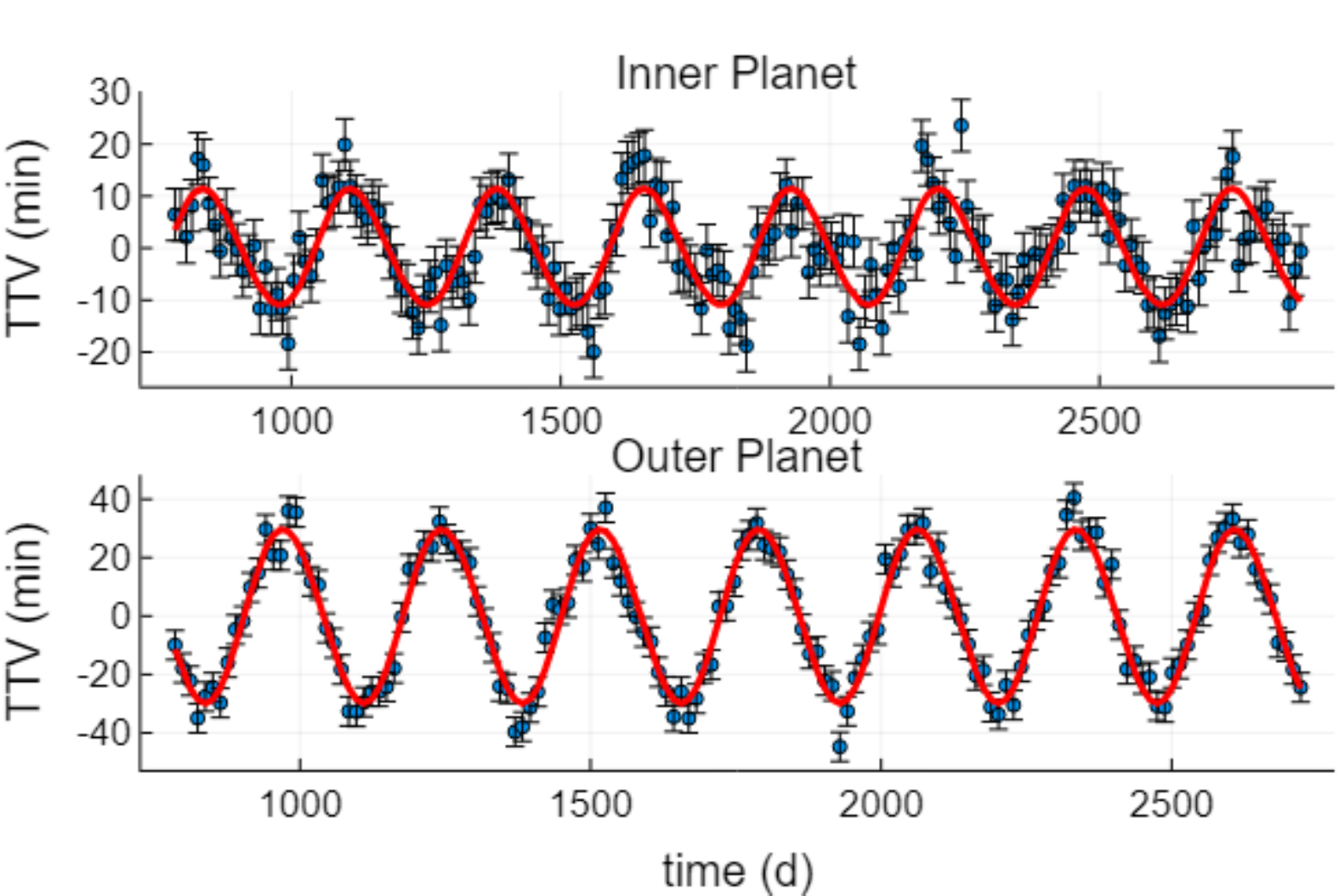}
\caption{Simulated dataset from the simple sinusoidal model (SSM) with 5 min of added Gaussian white noise. Model with true parameters shown in red.}
\label{SSMdata}
\end{figure}

To test samplers using this model, we simulated a dataset that was comparable to actual TTV data (see Fig  \ref{SSMdata}). Using recorded transit time data for Kepler-307, a well-behaved TTV system with two planets close to a 5:4 resonance, we compute the best fitting SSM parameters for the data \citep{rowe2014}. We generated a synthetic dataset for the SSM model using ``true'' parameters listed in Table \ref{SSMRecover}, adding 5 minutes of Gaussian white noise to each observation. For each of the model parameters, we used uniform priors for simplicity. In Section \ref{sec:SSMresults}, we explore which of the various MCMC samplers are able to sample efficiently and accurately from the resulting posterior distribution.

\subsubsection{TTVFaster Model}

The TTVFaster model uses a series expansion around orbital resonances in order to estimate TTVs for a given set of parameters. For both planets in the system, this model uses their planet-star mass ratios $\mu$, orbital periods $P$, initial transit times $t_i$, and eccentricity vector components $k = e \cos{\omega}$ and $h = e \sin{\omega}$ (where \textit{e} is eccentricity and $\omega$ is the argument of periastron). With an input of these 10 model parameters $p_i$ to the TTVFaster model, we obtain outputs of $\tau_m (N_m, p)$ for the transit times of planets 1 and 2 at transit numbers $N_m$. In order to obtain first and second order derivatives of the TTVFaster model, we modified the TTVFaster code\footnote{\url{https://github.com/nwtuchow/TTVFaster.git}} to work in the framework of forward automatic differentiation using the \texttt{ForwardDiff.jl} package \citep{forwarddiff}.
 %do we need to cite forwarddiff twice?

  We specified priors for TTVFaster model parameters to ensure that MCMC chains will only accept proposals for physically plausible parameter values. Priors for masses and periods were uniform and bounded to be positive. We specify that the period for the inner planet must be shorter than that of the outer planet, and required orbital eccentricities to be less than 1. We used nearly uniform priors in initial transit times, constant to within half a period, and smoothed the edges by steep Gaussians in order to avoid discontinuities in the prior or its derivatives. Gaussian priors with mean 0.0 and standard deviation 0.1 were used for eccentricity vector components $h$ and $k$, resulting in the eccentricity, $e$, being drawn from a Rayleigh distribution.
%write equations for priors?

\begin{figure}
\includegraphics[width=\columnwidth]{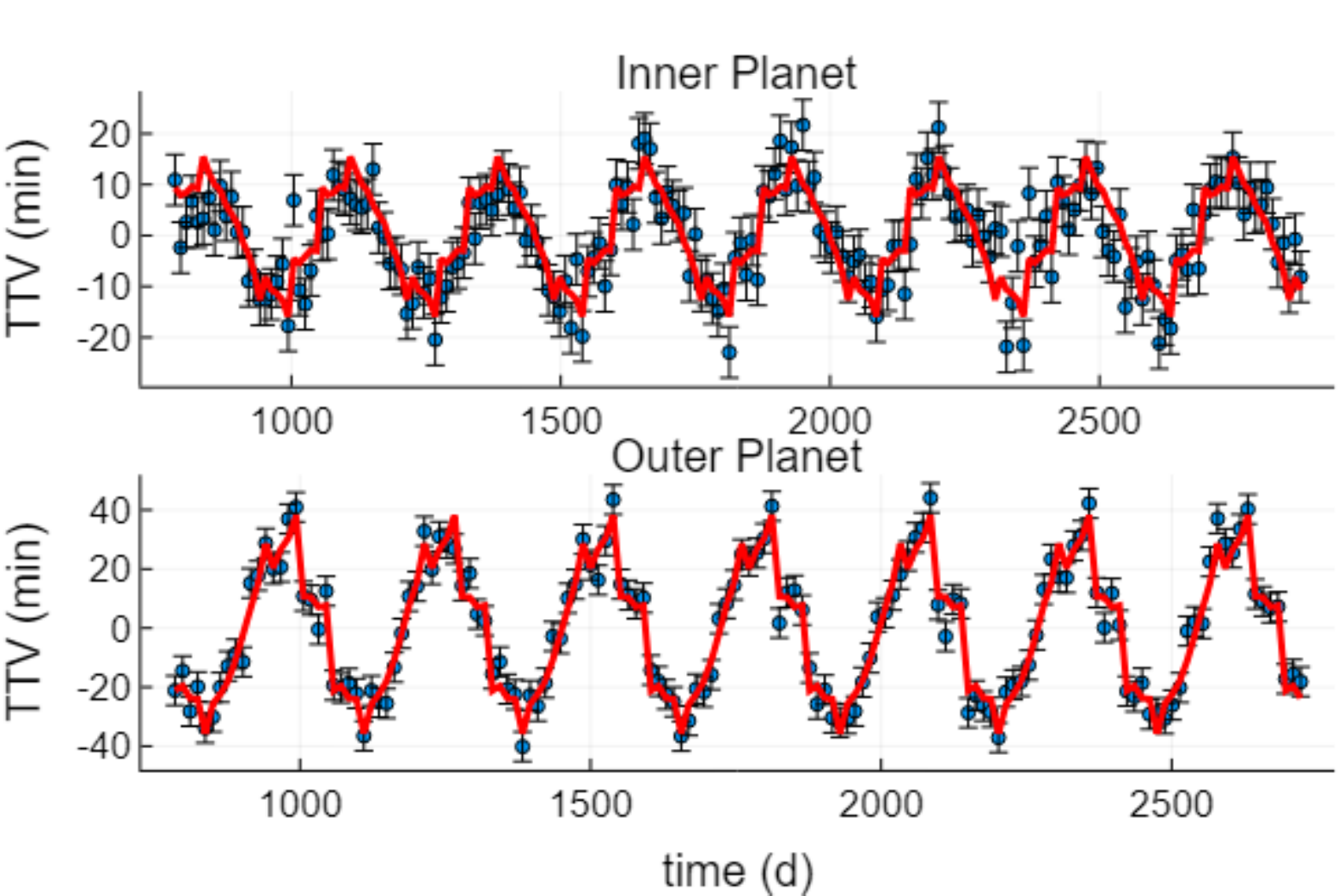}
\caption{Simulated dataset from TTVFaster based on the Kepler-307 (KOI-1576) system with 5 min of added Gaussian white noise. Model with true parameters shown in red.}
\label{k307data}
\end{figure}

We generated multiple synthetic datasets using TTVFaster, based on actual TTV systems of varying complexity. Previous studies have characterized the posterior distributions of multiple TTV systems using more computationally expensive N-body models, so we chose systems based on the apparent complexity of their posteriors computed in \citet{jontofhutter2016}. First, we considered the Kepler-307 (KOI-1576) system, a two planet system near a 5:4 mean motion resonance. We generated synthetic datasets using TTVFaster with true parameters close to those of Kepler-307 (see Fig \ref{k307data}), and simulated the times of transits which were observed in the  \citet{rowe2014} dataset.  For this system, we made two datasets, one adding 5 minutes of Gaussian white noise to the data, close to the actual measurement uncertainty, and one with 15 minutes of Gaussian white noise.

%modified: gave amount of noise
\begin{figure}
\includegraphics[width=\columnwidth]{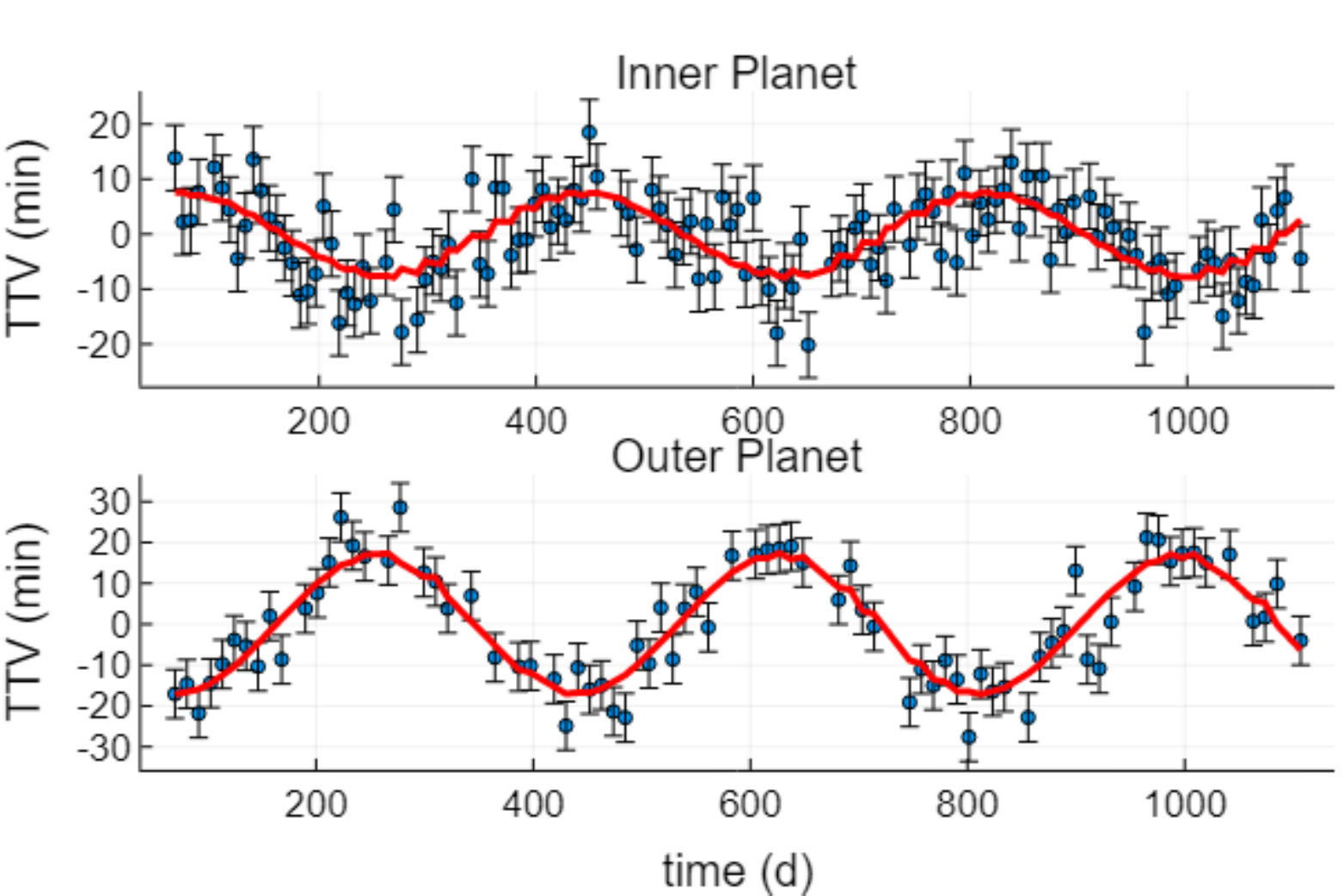}
\caption{Simulated dataset from TTVFaster based on the Kepler-49 (KOI-248) system with 5.96 minutes of added Gaussian white noise. Model with true parameters shown in red.}
\label{k49data}
\end{figure}

We also generated a synthetic dataset using TTVFaster with true parameters inspired by the slightly more challenging Kepler-49 (KOI-248) system (see Fig \ref{k49data}). The Kepler-49 system has 2 inner planets near a 3:2 orbital resonance, but also 2 outer planets that may slightly perturb the inner ones. In generating this dataset, we considered only the inner two planets, and simulated their interactions in the absence of the outer two. Previous studies showed that the masses and eccentricities for planets in this system were less well constrained than those of Kepler-307. We added 5.96 minutes of Gaussian white noise to these data, using the mean uncertainties for transit times in the observed dataset.

%modified: gave amount of noise
\begin{figure}
\includegraphics[width=\columnwidth]{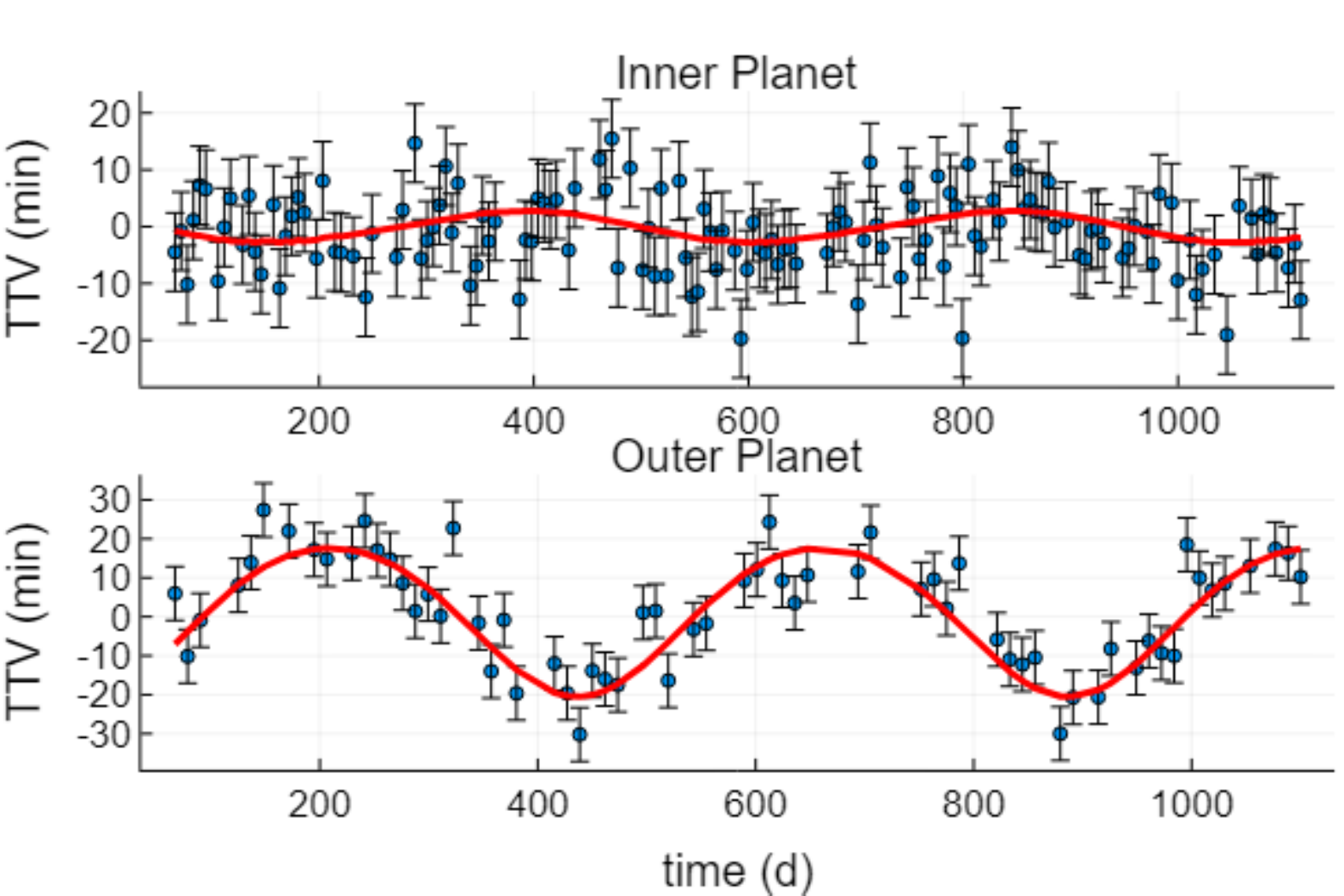}
\caption{Simulated dataset from TTVFaster based on the Kepler-57 (KOI-1270) system with 6.90 minutes of added Gaussian white noise. Model with true parameters shown in red.}
\label{k57data}
\end{figure}

Finally we generate a synthetic dataset using true parameters inspired by those of the Kepler-57 (KOI-1270) system (see Fig \ref{k57data}). Previous studies had difficulty constraining the planet masses and the posteriors for eccentricity h and k components showed bimodality. We added Gaussian white noise to the synthetic data, with a standard deviation of 6.90 minutes given by the average uncertainty in the observed data. These datasets were used to test the efficiencies and accuracies of the MCMC samplers under consideration.

\subsection{Sampling and Tuning}

We compared the performance of six types of samplers: the MALA, HMC , SMMALA, GAMC \footnote{With an exponential decay update schedule (r=1/50000)}, DEMCMC, and AIMCMC  samplers described in Section \ref{sec:statbackground}.
To sample from the posterior distributions for each model, we used the framework and MCMC samplers provided within the \texttt{Klara.jl} package\footnote{\url{https://github.com/JuliaStats/Klara.jl.git}}, our \texttt{TTVMCMC} repository\footnote{\url{https://github.com/nwtuchow/TTVMCMC.git}}, as well as the GAMC sampler in \texttt{GAMCSampler.jl}\footnote{\url{https://github.com/scidom/GAMCSampler.jl.git}}.  While many of the samplers we tested are ideal for running in parallel, to fairly compare the efficiencies of the different samplers, we focused on assessing their performance running on the same machine without the use of multiple cores.

As we were primarily concerned with the performance of the samplers near the posterior maxima, the starting points for the MCMC chains for each model were chosen to be close to the parameter values used to generate the datasets. Parameters such as periods, which dictate the frequency of the sinusoidal waveform in either class of model, needed to start close to their actual values from the datasets, or the MCMC chains would get stuck at local maxima near multiples of the ``true'' periods. Terms determining the amplitude of the waveform were set to start farther from their ``true'' values, choosing initial values that could reasonably be several standard deviations away from the target values. Starting at these locations, we ran pilot runs to burn-in each of the MCMC chains and find the posterior maxima (see Section \ref{sec:transform}). For our diagnostics of sampler efficiency, we initialize chains at the position of the last iteration of the pilot run to ensure that we are measuring the sampling efficiency near the mode of the posterior

\begin{figure}
\includegraphics[width=\columnwidth]{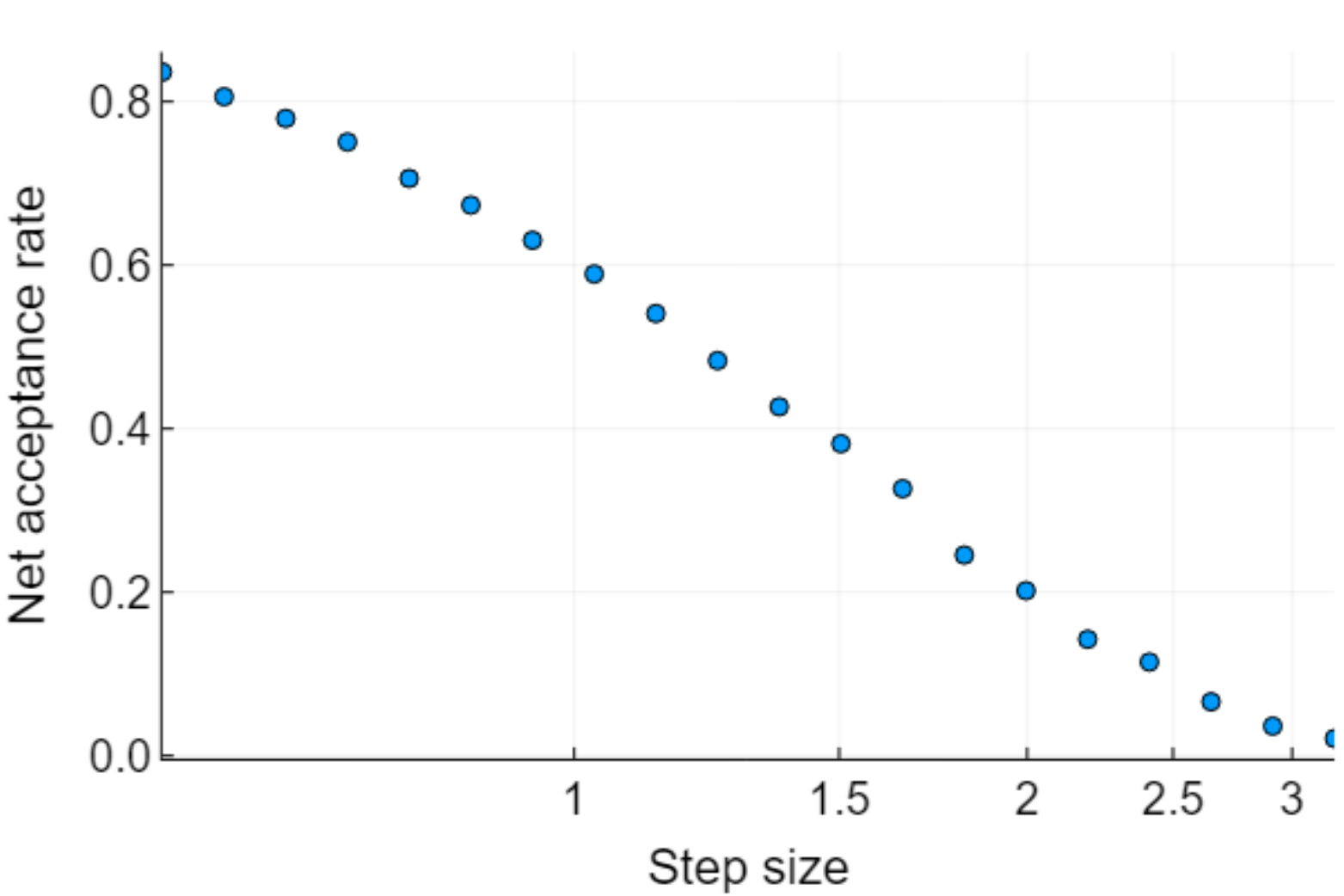}
\caption{Acceptance rate as a function of MALA step size tuning parameters. This example used 5,000 iteration pilot MCMC chains on the Kepler-307 (5 min) TTVFaster model.}
\label{acceptPlot}
\end{figure}

\begin{figure}
\includegraphics[width=\columnwidth]{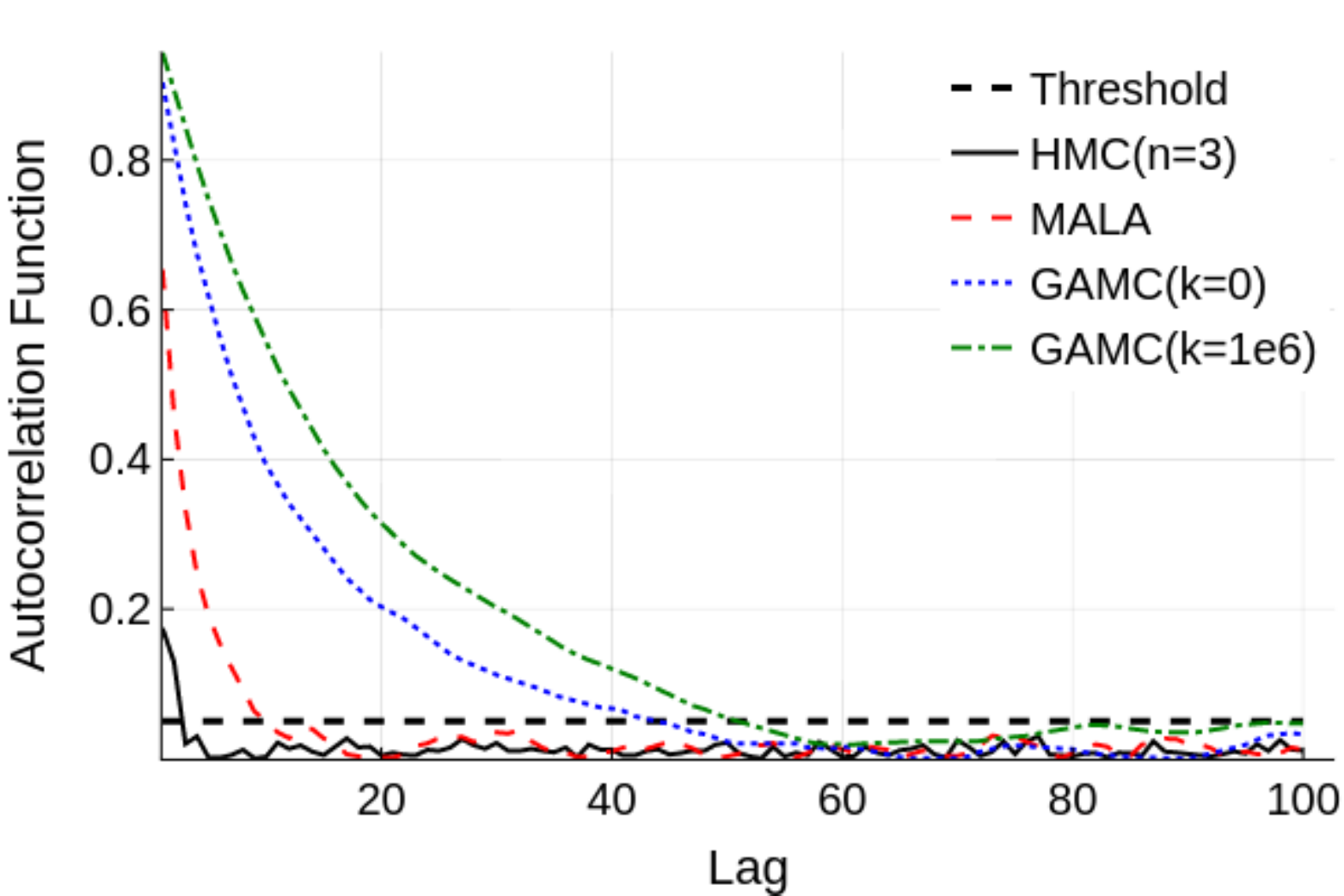}
\caption{Autocorrelation Functions for parameter $\mu_1$ for the Kepler-307 inspired TTVFaster model. Absolute values of autocorrelation functions are shown here for MCMC chains from different tuned samplers. The autocorrelation length is defined as the intercept between the autocorrelation function and a user defined threshold (in this case 0.05). Autocorrelation lengths for $\mu_1$ for the HMC($n=3$), MALA, GAMC($k=0$) and GAMC($k=10^6$) samplers are [3,11,45,51] respectively.}
\label{autocorrFunc}
\end{figure}

Before determining which sampler yielded the most efficient sampling, we first needed to tune the key algorithmic parameter of each MCMC sampler to improve their efficiencies. We used several diagnostics to assess the efficiency when tuning the MCMC samplers. The simplest of these was to calculate the acceptance rates for pilot chains. The step size should be small enough that a substantial fraction of the proposed steps are accepted, but large enough that the proposals are significantly different, and do not only propose states which are very close to the current chain position.  Fig \ref{acceptPlot} shows an example of the acceptance rate versus the drift step size tuning parameter for the MALA sampler. Additionally, we calculated the autocorrelation length, $\lambda$, for each MCMC chain, which we define as the smallest value of lag required for the autocorrelation function to pass below a user defined threshold of 0.05 (see Fig \ref{autocorrFunc}).  We calculate the effective sample size (ESS) for each parameter by dividing the MCMC chain length by the autocorrelation length.

Our tuning process was as follows. Initially, we found values of the tuning parameters where acceptance rates were between $90\%$ and $10\%$, and generated a logarithmically spaced array of tuning parameter values between these two extremes. We ran MCMC pilot chains (varying in length depending on the model) for each proposed sampler tuning parameter value, and recorded the acceptance rate as well as autocorrelation lengths ($\lambda$) and effective sample sizes. For a given MCMC chain, each model parameter has its own $\lambda$ and ESS value, and we define $\lambda_{max}$ as the maximum of the $\lambda$ values for the different model parameters. For each tuning parameter value, we compute $\lambda_{max}$  and set the tuning parameter to be the value that minimizes $\lambda_{max}$.
%rephrase
After adjusting the sampler tuning parameters, we ran longer MCMC chains to sample from the underlying target distribution. We assessed whether the samplers were effectively exploring parameter space using a diagnostic of minimum ESS over elapsed wall clock time.

\subsection{Coordinate Transform and Pilot Run}
\label{sec:transform}

One of the problems we encountered when sampling from the posteriors of these models is that the different parameters typically have values that vary by orders of magnitude.  For example, planet-star mass ratios have magnitudes of order $10^{-5}$, but initial transit times take on values spanning roughly $10^2$ days. This makes it difficult for samplers to propose reasonably sized steps to explore the posterior distribution efficiently.
 Another challenge is that posteriors for certain parameters, such as masses and eccentricities, are often highly correlated with each other, resulting in much slower convergence of Markov chains. Finally, some samplers, such as GAMC, require computing the Cholesky decomposition of the inverse Hessian, which can result in positive definite errors due to round-off errors in finite precision arithmetic.

To resolve all of these issues, we use a linear coordinate transform to both rotate and scale the parameter space. In many cases, a linear transformation can often result in a nearly uncorrelated parameter space with magnitudes of order unity.
%motivation for using this transform
Therefore, we sample from a space of transformed parameters, $z$, related to model parameters by
\begin{equation}
z = \mathbf{\tilde \Sigma}^{-1/2} (p- \tilde p_{means})
\end{equation}
where $\mathbf{\tilde \Sigma}$ and $\tilde p_{means}$ are estimates of the sample covariance matrix and parameter means respectively. (Note that for $\mathbf{\tilde \Sigma}^{-1/2}$  we take the lower triangular for the Cholesky decomposition).

Initially, we compute an estimate of the covariance matrix by taking the inverse of the negative Hessian matrix at the chain starting position and use the starting position of the chain as a rough estimate of the means.
Applying these estimates in the transform, we then run a 500,000 iteration burn-in chain with the GAMC sampler.
We update our estimates of both posterior covariances and means by computing the sample covariances and means of this longer pilot chain. This process can be repeated multiple times to get progressively better estimates of covariances and means.

To burn-in ensemble samplers like DEMCMC and AIMCMC, we used a slightly different pilot run. We generated random starting points for each walker, drawing from a multivariate Gaussian with sample covariances and means calculated from the earlier single walker pilot chains. We ran the samplers for 50,000 generations to burn-in the chains for each walker in the ensemble prior to computing autocorrelation lengths and effective sample sizes.
%how did we tell if estimates were good

\section{Results}

For each model, we tested the DEMCMC, AIMCMC, MALA, HMC, SMMALA, and GAMC samplers to evaluate how efficiently each sampled from the posterior distribution. Many samplers, such as MALA and SMMALA, only have a single step size tuning parameter, so their performance can be assessed after simply tuning the one parameter. The HMC sampler has two main tuning parameters: the leapfrog step size and number of steps, $n$, in the proposed chain. We tested efficiencies for different fixed numbers of leapfrog steps, tuning the step size parameter separately for each $n$.
Similarly, the GAMC sampler has tuning parameters for the drift step size, as well as one for the exponential schedule used (see Appendix \ref{appendix:GAMC}). We tuned the GAMC step size parameter for update schedules starting at fixed numbers of iterations, $k$.
%change symbol for k to something unused elsewhere in the paper?
For ensemble samplers like DEMCMC and AIMCMC, the performance is insensitive to any step size scale parameter, so we assessed the performance with the default scale parameter.
%cite?

To compare the efficiencies of the different samplers, we used two main measures of sampling efficiency. For each tuned sampler, we compute pilot chains for 10,000 iterations, starting at the position of the last iteration of the pilot runs used to burn-in the chains, and measure their elapsed run times. We diagnosed the efficiency of samplers using the mean effective sample size (ESS) divided by elapsed wall clock time, or the minimum effective sample size divided by wall clock time\footnote{Means and minimums here are over ESS values for each parameter}.
%should I only report min(ess)/time?
Using the sampler with the highest efficiency, we proceeded to run longer MCMC chains for computing the posterior distributions.  Since we are using synthetic datasets, we test whether the credible intervals of the posterior distributions include the true parameter values used to generate the dataset.

%modified, adding AIMCMC
\begin{table*}
\caption{Diagnostics for sampler efficiencies for TTV Models. ESS refers to effective sample size. Elapsed Times are for 10,000 iteration runs.}
\label{DiagnosticTable}
\begin{tabular}{lcccc}
\hline
\textbf{Sampler} & \textbf{Step size} & \textbf{Elapsed Time (s)} & \textbf{Mean(ESS)/time (s$^{-1}$)} & \textbf{Min(ESS)/time (s$^{-1}$)} \\ \hline
\multicolumn{5}{c}{\textit{Simple Sinusoidal Model}}\\ \hline
HMC($n=1$)          & 1.03  & 7.8  & 249. & 214. \\
HMC($n=2$)          & 0.918 & 13.3 & 750. & 750.  \\
HMC($n=3$)          & 0.736 & 19.1 & 167. & 131. \\
HMC($n=5$)            & 0.822 & 31.6 & 158. & 158. \\
HMC($n=7$)           & 1.15  & 40.6 & 246. & 246. \\
MALA            & 1.10   & 7.6 & 263. & 263. \\
SMMALA          & 0.953 & 79.9 & 24.3 & 20.9   \\
GAMC($k=0$)     & 1.15  & 56.7 & 12.3 & 9.80  \\
GAMC($k=25000$) & 0.527 & 3.3 & 74.8  & 52.9  \\
GAMC($k=50000$) & 1.03  & 2.6    & 67.8  & 43.3  \\
GAMC($k=10^6$)   & 0.736 & 2.6  & 85.7  & 56.8  \\
DEMCMC          & ---   & 60.1 & 113. & 109. \\
AIMCMC	       & ---  &	95.1	&21.5 &	20.4 \\
\hline
\multicolumn{5}{c}{\textit{Kepler-307 (5 min) model}}\\ \hline
HMC($n=1$)            & 0.918 & 52.6   & 27.5   & 19.0  \\
HMC($n=2$)            & 0.822 & 88.5   & 71.6   & 37.7   \\
HMC($n=3$)            & 0.717 & 119.8   & 27.8  & 27.8  \\
HMC($n=5$)            & 0.736 & 187.4 & 13.1  & 10.7   \\
HMC($n=7$)            & 1.03  & 254.3  & 6.56    & 4.92   \\
MALA                      & 0.974 & 56.8   & 28.3   & 25.2  \\
SMMALA                 & 0.464 & 2873.5 & 0.205 & 0.134 \\
GAMC($k=0$)        & 0.88  & 1824.7 & 0.235 & 0.157\\
GAMC($k=25000$) & 0.88  & 37.2   & 4.46   & 3.12  \\
GAMC($k=50000$) & 0.88  & 22.6   & 10.8  & 7.14   \\
GAMC($k=10^6$)   & 0.88  & 23.0   & 11.3  & 9.45   \\
DEMCMC          & ---   & 614.6  & 10.3  & 9.07   \\ 
AIMCMC	       & ---    &640.5	  &3.06   &	2.76 \\
\hline
\multicolumn{5}{c}{\textit{Kepler-307 (15 min) model}}\\ \hline
HMC($n=1$)            & 1.15  & 51.6   & 15.3  & 6.46 \\
HMC($n=2$)            & 0.838 & 81.0   & 25.8  & 2.17  \\
HMC($n=3$)            & 0.755 & 113.9  & 66.4  & 9.75 \\
HMC($n=5$)            & 0.838 & 156.6  & 8.42  & 2.78 \\
HMC($n=7$)            & 0.755 & 222.1  & 3.25  & 2.05 \\
MALA            & 0.550  & 51.2  & 16.2  & 10.9 \\
SMMALA          & 0.755 & 2943.7  & 0.224 & 0.110 \\
GAMC($k=0$)     & 0.402 & 1841.7 & 0.187 & 0.132 \\
GAMC($k=25000$) & 0.550  & 41.6    & 5.01   & 3.82  \\
GAMC($k=50000$) & 0.612 & 22.4   & 7.55   & 3.31  \\
GAMC($k=10^6$)   & 1.42  & 22.4   & 8.50   & 6.20  \\
DEMCMC          & ---   & 609.4  & 8.93    & 6.47   \\ 
AIMCMC	       & ---    & 632.7  &	2.89	& 2.27  \\
\hline
\end{tabular}
\end{table*}

%modified adding AIMCMC
\begin{table*}
\contcaption{Diagnostics for sampler efficiencies for TTV Models. ESS refers to effective sample size. Elapsed Times are for 10,000 iteration runs.}
\label{DiagnosticTableCont}
\begin{tabular}{lcccc}
\hline
\textbf{Sampler} & \textbf{Step size} & \textbf{Elapsed Time (s)} & \textbf{Mean(ESS)/time (s$^{-1}$)} & \textbf{Min(ESS)/time (s$^{-1}$)} \\ \hline
\multicolumn{5}{c}{\textit{Kepler-49 model}}\\ \hline
HMC($n=1$)            & 0.0544  & 29.7   & 0.387  & 0.211  \\
HMC($n=2$)            & 0.0584  & 47.1   & 0.188 & 0.0872 \\
HMC($n=3$)            & 0.0421  & 72.1   & 0.236  & 0.0610 \\
HMC($n=5$)            & 0.049   & 111.5   & 0.519  & 0.290  \\
HMC($n=7$)            & 0.0397  & 153.7  & 0.873  & 0.626  \\
MALA            & 0.00538 & 34.6   & 0.229 & 0.104  \\
SMMALA          & 0.283   & 1984.0 & 0.137 & 0.0118 \\
GAMC($k=0$)     & 0.868   & 1160.3  & 0.218 & 0.00917 \\
GAMC($k=25000$) & 0.14    & 30.2    & 3.05   & 0.265 \\
GAMC($k=50000$) & 0.349   & 12.7   & 8.04  & 3.64  \\
GAMC($k=10^6$)   & 0.349   & 12.4   & 9.49    & 2.48  \\
DEMCMC          & ---     & 270.9  & 2.82   & 1.34  \\
AIMCMC	        & ---	 &310.8   &1.93    & 0.900 \\

 \hline
\multicolumn{5}{c}{\textit{Kepler-57 model}}\\ \hline
HMC($n=1$)            & 0.402  & 20.2   & 1.11  & 0.351  \\
HMC($n=2$)            & 0.173  & 34.8   & 2.24   & 0.943  \\
HMC($n=3$)            & 0.14   & 48.2   & 4.09   & 1.23   \\
HMC($n=5$)            & 0.126  & 75.1   & 5.66  & 2.05   \\
HMC($n=7$)            & 0.192  & 90.1   & 5.39   & 1.32    \\
MALA            & 0.0605 & 20.7   & 1.75  & 0.452   \\
SMMALA          & 0.126  & 1915.5 & 0.0754 & 0.00838 \\
GAMC($k=0$)     & 0.402  & 1158.0 & 0.198 & 0.0332  \\
GAMC($k=25000$) & 0.838  & 22.8   & 7.06   & 2.41    \\
GAMC($k=50000$) & 0.612  & 8.5    & 15.4  & 4.62    \\
GAMC($k=10^6$)   & 0.612  & 8.6    & 17.6  & 0.835   \\
DEMCMC          & ---    & 185.1  & 10.2  & 4.55    \\ 
AIMCMC	        & ---    & 203.9   &4.82  &2.61     \\
\hline
\end{tabular}
\end{table*}

\subsection{Simple Sinusoidal Model}
\label{sec:SSMresults}

Table \ref{DiagnosticTable} presents the results of the efficiency diagnostics for the SSM model, along with the elapsed times for 10,000 iteration MCMC chains and the tuned drift step size parameters (or equivalent) for each sampler. The HMC sampler with $n=2$ leapfrog steps scores the best, with a minimum ESS per time that is significantly greater than those of the other samplers. Nevertheless, many of the other samplers, such as HMC with different numbers of steps and MALA,  perform well and obtain large ESSs in relatively short elapsed times.
%It is a testament to both the efficacy of the coordinate transform applied and the simplicity of the SSM model that we obtain such large values for ESS per time, allowing us to obtain

We observe that the samplers that use the Hessian of the target distribution take longer to run than the samplers which only use the target gradient. The SMMALA sampler, which uses the Hessian at every step, takes about 10 times as long per step as the MALA sampler, but both algorithms obtain comparable ESS's, resulting in SMMALA obtaining a much lower value for Min(ESS)/time. The GAMC sampler uses the Hessian on an exponentially decaying schedule, so, at the zero-th MCMC iteration ($k=0$), it performs comparably to SMMALA. However, its speed dramatically increases at later iterations, when it primarily takes adaptive Metropolis (AM) steps and Hessian updates are much less frequent. After many iterations, the GAMC sampler runs faster than any of the gradient based samplers, but there is a trade off in effective sample size, as the AM steps are typically more correlated with each other than the SMMALA steps. We can see that, among the ensemble samplers, DEMCMC performs much more efficiently than AIMCMC, but still performs worse than the samplers which use the target gradient.

%modified caption, specifying units
\begin{table}
\caption{``True'' values used to generate the synthetic dataset and summary statistics for SSM model parameter values. All parameters are in units of days.}
\label{SSMRecover}
\begin{tabular}{lll}
\hline
\textbf{Parameter} & \textbf{True Value} & \textbf{Recovered}
\\ \hline
$t_{i,1}$                 &783.9985        & $783.9981 \pm 0.0005$
\\
$P_1$                    &10.500011             & $10.500013 \pm 0.000004$
\\
$A_1$                   &  0.0034               & $0.0036 \pm 0.0003$
\\
$B_1$                   &  0.0070         & $0.0064 \pm 0.0003$
\\
$C_1$                  &$-5.87 * 10^{-5}$  &  $(-3.2 \pm 3.4) *10^{-4}$
\\
$D_1$                     & 0.0004        & $0.0011 \pm 0.0003$
\\
$t_{i,2}$                 & 785.000061          & $784.9996 \pm 0.0006$
\\
$P_2$                     &  13.000001           & $13.000006 \pm 0.000007$
\\
$A_2$                    & -0.0077                & $-0.0075 \pm 0.0004$
\\
$B_2$                          & -0.0192                & $-0.0195 \pm 0.0004$
\\
$C_2$                          & -0.000337754    & $-0.0007 \pm 0.0004$
\\
$D_2$                          &  0.000220712     & $-0.00007 \pm 0.00040$
\\ \hline
\end{tabular}
\end{table}

Using the HMC sampler with $n=2$ leapfrog steps, we ran a 2,000,000 iteration MCMC chain to sample from the posterior distribution for the SSM.  In the resulting chain, the 2D marginal posteriors appear as nearly Gaussian ellipsoids, and only the parameters for periods and times of first transit appear slightly correlated ($t_{i,1}$ vs $P_1$ and $t_{i_,2}$ vs. $P_2$). This smooth, nearly Gaussian posterior explains why many samplers resulted in such large Min(ESS)/time. In Table \ref{SSMRecover}, we compare the true values of parameters used to generate the dataset and the median values for parameters from the posterior distribution. For the times of first transits, periods, and outer planet's parameters, the true values are very close to those recovered from the posterior, falling within the central 68\% credible interval (CI) of the median values.
However, for parameters $B_1$ and $D_1$, the true values don't fall within 68\% credible interval of the recovered values. This is not unexpected, as the noise added to the dataset changes which exact parameter values maximize the likelihood, and the inner planet has lower signal to noise than the outer planet.

\subsection{TTVFaster Model}
\subsubsection{Kepler-307 (KOI-1576) Datasets}

We analyzed two synthetic datasets inspired by observations of the well-behaved Kepler-307 system: one with 5 minutes of Gaussian white noise (similar to the uncertainty in actual datasets), and one with 15 minutes of Gaussian white noise. For the 5 minute noise model, the results of our diagnostics can be found in Table \ref{DiagnosticTable}. We can see that, similarly to the SSM model, the HMC ($n=2$) sampler performs the best in terms of Min(ESS)/time. Other samplers such as MALA and HMC at $n=1$ and $n=3$ steps also appear to perform almost as well. We notice that, for all the samplers, these 10,000 iteration pilot chains had significantly longer run times than those for the SSM, because taking the gradient and Hessian of the TTVFaster model is much more computationally demanding. This is evident by the fact that the values for Min(ESS)/time are all significantly less than those for the SSM by about an order of magnitude or more. One can observe that the samplers which use the target Hessian the most, the SMMALA and GAMC ($k=0$) samplers, take roughly 10 times longer to run than the gradient using samplers, and don't result in a significantly larger ESS. This causes them to perform the worst, and they are likely better suited for sampling from more challenging posteriors.

For the Kepler-307 inspired dataset with 15 minutes of Gaussian white noise, the results of our diagnostics can be seen in Table \ref{DiagnosticTable}. The elapsed times are comparable to those of the 5 minute noise dataset, but effective sample sizes are somewhat diminished. We find that the MALA and HMC ($n=3$) samplers achieve the highest sampling efficiency, followed by samplers such as HMC ($n=1$), DEMCMC, and GAMC after one million iterations.  For both datasets, we observe that the AIMCMC sampler again has lower efficiency that DEMCMC.

\begin{figure*}
\includegraphics[width=\textwidth]{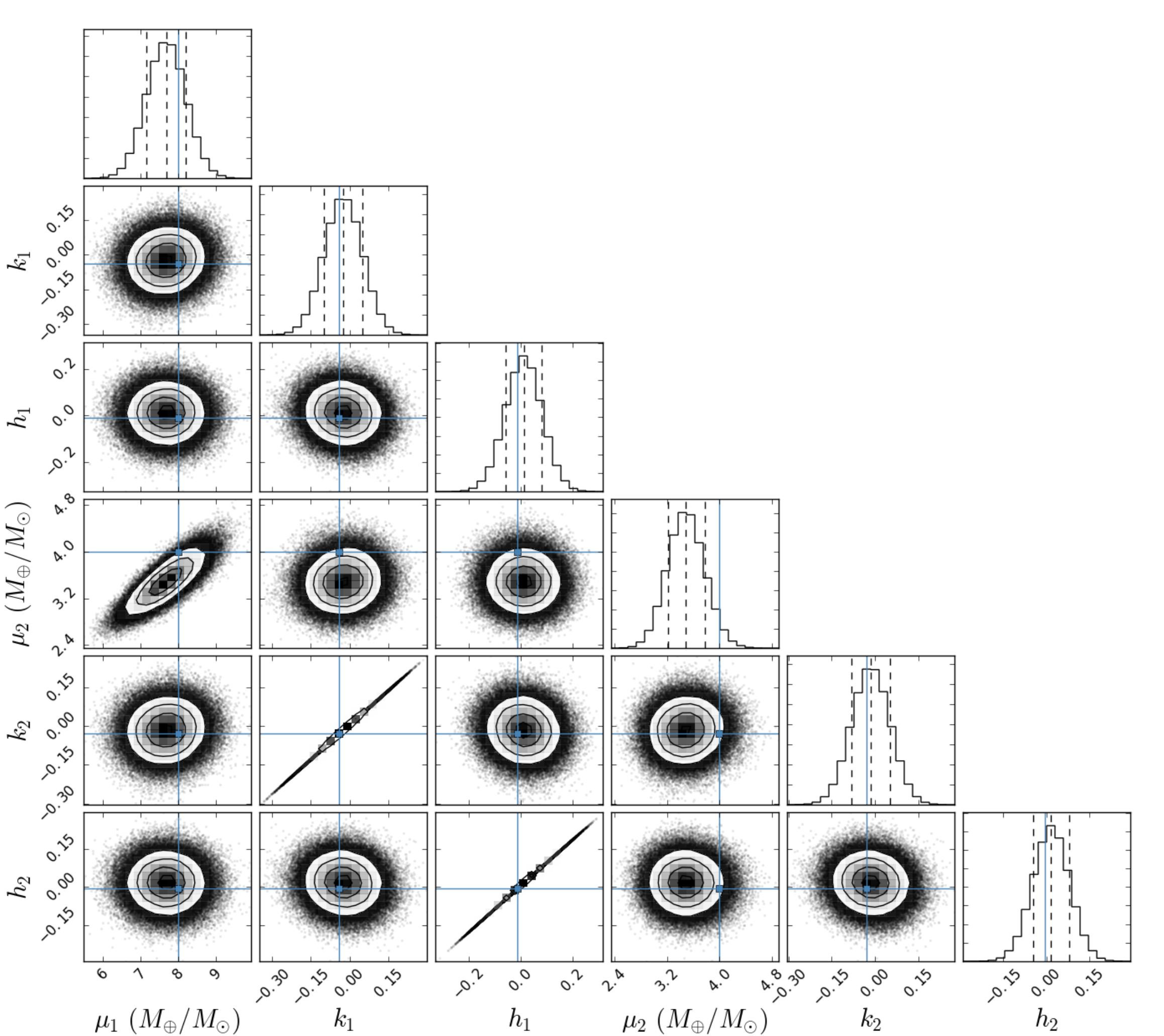}
\caption{Corner plot showing the posterior distribution for the Kepler-307 (KOI-1576) TTVFaster model with 5 minutes of noise. This chain used the tuned HMC ($n=2$) sampler on a transformed parameter space and ran for 2 million iterations with a 500,000 iteration burn-in period. Figure created using the \texttt{corner.py} package \citep{corner}.}
\label{k307Corner}
\end{figure*}

%modified specifying units
\begin{table}
\caption{``True'' values used to generate the Kepler-307 (5 min noise) inspired dataset and summary statistics from the posterior distribution in Figure \ref{k307Corner}. Mass ratios and eccentricities reported here are unitless.}
\label{k307Recover}
\begin{tabular}{lll}
\hline
\textbf{Parameter} & \textbf{True Value} & \textbf{Recovered}
\\ \hline
$\mu_1$               & $2.4024 * 10^{-5}$  &  $(2.31 ^{+0.15} _{-0.16}) * 10^{-5}$
\\
$P_1$ (d)                  & 10.500000                 & $10.500011 \pm 0.000004$
\\
$t_{i,1}$ (d)             & 784.0000                 & $783.9985 \pm 0.0005$
\\
$k_1$                   & -0.040                      &  $-0.025 \pm 0.073$
\\
$h_1$                   &  -0.011                     & $0.012 ^{+0.069} _{-0.070}$
\\
$\mu_2$               & $1.2012 * 10^{-5}$  & $ (1.05 ^{+0.09} _{-0.08})* 10^{-5}$
\\
$P_2$ (d)                  & 13.000000                  & $13.000000 ^{+0.000006} _{-0.000007}$
\\
$t_{i,2}$ (d)              & 785.0000                   & $785.0002 \pm 0.0006$
\\
$k_2$                    & -0.029                       & $-0.015 \pm 0.066$
\\
$h_2$                    & -0.004                      & $0.017 ^{+0.063} _{-0.064}$
\\ \hline
\end{tabular}

\end{table}

%modified specifying units
\begin{table}
\caption{``True'' values used to generate the noisier Kepler-307 (15 min noise) dataset and summary statistics from the posterior distribution. Mass ratios and eccentricities reported here are unitless.}
\label{NoisyRecover}
\begin{tabular}{lll}
\hline
\textbf{Parameter} & \textbf{True Value} & \textbf{Recovered}
\\ \hline
$\mu_1$               & $2.4024 * 10^{-5}$  &  $(2.94 \pm 0.48)*10^{-5}$
\\
$P_1$ (d)                  & 10.500000                 & $10.50000 \pm 0.00001$
\\
$t_{i,1}$ (d)             & 784.0000                   & $784.000 \pm 0.001$
\\
$k_1$                   & -0.040                      &  $-0.000 \pm 0.074$
\\
$h_1$                   &  -0.011                     & $0.003 \pm 0.073$
\\
$\mu_2$               & $1.2012 * 10^{-5}$  & $(1.23 ^{+0.26} _{-0.23}) * 10^{-5}$
\\
$P_2$ (d)                  & 13.000000                 & $12.99999 \pm 0.00002$
\\
$t_{i,2}$ (d)              & 785.0000                   & $785.000 \pm 0.002$
\\
$k_2$                    & -0.029                       & $0.005 \pm 0.066$
\\
$h_2$                    & -0.004                      & $0.009 \pm 0.066$
\\ \hline
\end{tabular}

\end{table}

For the 5 and 15 minute noise Kepler-307 datasets, we ran longer MCMC runs to converge on the posterior distributions. Figure \ref{k307Corner} shows the posterior distribution for the 5 minute Gaussian white noise model (omitting nearly Gaussian Periods and times of first transits). The posterior for the 15 minute Gaussian white noise model appears very similar, but with less constrained parameter values. For both datasets, the 2-D marginal posteriors for many of the parameters are smooth ellipsoids and have marginal distributions that are close to Gaussian. Besides the correlation between orbital periods and times of first transits, seen in the SSM, this TTVFaster model displays strong correlations between the eccentricity components, $h$ and $k$, of the two planets, as well as a weaker correlation between planet mass ratios. Tables \ref{k307Recover} and \ref{NoisyRecover} show the median values recovered from the posteriors of both models compared to the true values used to generate them.  Most of the true values for both models fall within the central 68\% CI of the recovered values, but for the 5 minute model, the  true value for $\mu_2$ is greater than the upper bound of the 68\% CI of the recovered value, and $t_{i,1}$ differs slightly by less than the amount of added noise. For the 15 minute noise model, the true value for $\mu_1$ falls just beneath the lower bound of the credible interval for the recovered value. Mass ratios for the inner planets are less well constrained than those of the outer planet in both models, likely due to the fact that the inner planet has lower signal to noise in both datasets. The uncertainties in $h$ and $k$ components are often larger than their magnitudes, resulting in weak constraints on the eccentricities, despite strong constraints on the ratios of $k_2 / k_1$ and $h_2/h_1$.

\subsubsection{Kepler-49 (KOI-248) Dataset}

We present the results of our diagnostics for a dataset generated by a TTVFaster model with true parameters similar to those of Kepler-49 system in Table \ref{DiagnosticTable}. The step sizes that maximize sampling efficiency for many samplers are much smaller here than for the earlier models for Kepler-307. Since the underlying distribution for this model is more irregular and challenging to sample from, the chains need to take smaller steps to reach reasonable acceptance rates and maximize effective sample sizes. According to the Min(ESS)/time diagnostic, GAMC performs the best after $k=50,000$ iterations. The reason that this diagnostic value is higher after 50,000 GAMC iterations than after one million is unclear, but perhaps the frequency of Hessian updates at $k=50,000$ allows the GAMC chains to sample most efficiently without sacrificing evaluation time. After $k=10^6$ iterations, the minimum ESS per time is lower than that of $k=50,000$, but the mean ESS per time is higher, suggesting that one parameter may get stuck in a  region of parameter space, and that Hessian evaluations help to get parameters unstuck.
%talk about how its not clear reason for efficiency peak
%note value of r*k is what matters
%but for k  -> k +10,000 iteration chains, r used is important
The DEMCMC sampler was almost as efficient as the GAMC sampler, while SMMALA and GAMC (k=0) display the worst efficiency due to the frequent computationally expensive Hessian evaluations. Here we can again see AIMCMC trails behind DEMCMC in efficiency, but performs better than many of the other samplers. The samplers using only target gradients, HMC and MALA, also don't perform as efficiently as they did for earlier models.

\begin{figure*}
\centering
\includegraphics[width=\textwidth]{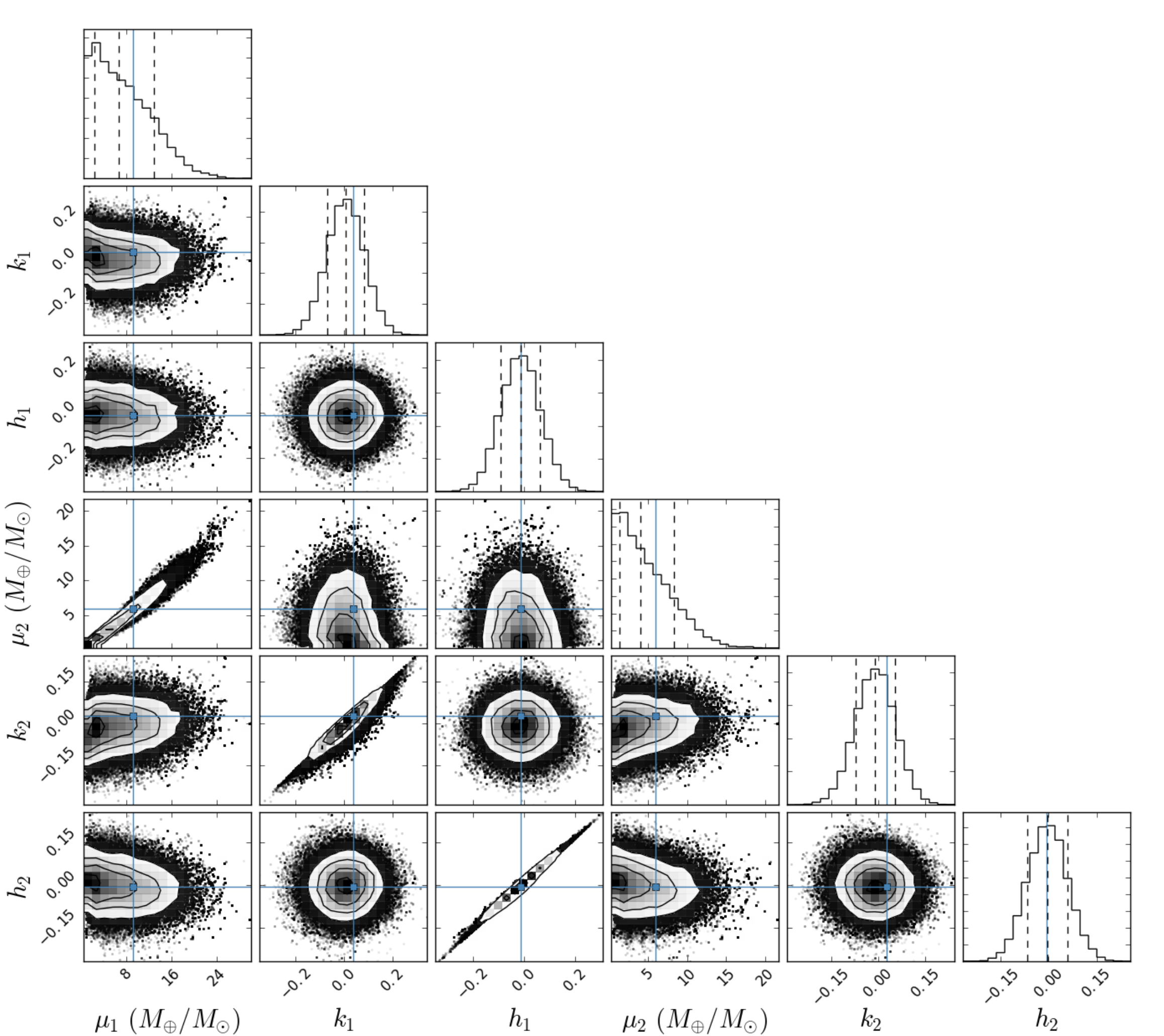}
\caption{Corner plot showing the posterior distribution for the Kepler-49 (KOI-248) TTVFaster model. This chain used the tuned GAMC sampler on a transformed parameter space and ran for 5 million iterations with a 500,000 iteration burn-in period.}
\label{k49Corner}
\end{figure*}

%modified specifying units
\begin{table}
\caption{``True'' values used to generate the Kepler-49 dataset and summary statistics from the posterior distribution in Figure \ref{k49Corner}. Mass ratios and eccentricities reported here are unitless.}
\label{k49Recover}
\begin{tabular}{lll}
\hline
\textbf{Parameter} & \textbf{True Value} & \textbf{Recovered}
\\ \hline
$\mu_1$               & $2.75 * 10^{-5}$  &  $(2.0 ^{+1.9} _{-1.3}) *10^{-5}$
\\
$P_1$ (d)                  & 7.204000                   & $7.204005 \pm 0.000009$
\\
$t_{i,1}$ (d)             &  780.4530                   & $780.4526 \pm 0.0004$
\\
$k_1$                   &  0.037                     &  $0.006 \pm 0.077$
\\
$h_1$                   &  -0.011                    & $-0.013 \pm 0.077$
\\
$\mu_2$               & $1.77 * 10^{-5}$  & $(1.2 ^{+1.3} _{-0.8})*10^{-5}$
\\
$P_2$ (d)                  & 10.9123                      & $10.91228 \pm 0.00002$
\\
$t_{i,2}$ (d)              & 790.347                      & $790.3468 \pm 0.0006$
\\
$k_2$                    & 0.027                       & $-0.010 ^{+0.063} _{-0.064}$
\\
$h_2$                    &  -0.006                      & $-0.003 ^{+0.062} _{-0.063}$
\\ \hline
\end{tabular}

\end{table}

 The reasons for our decreased sampling efficiencies for some samplers can be seen in Figure \ref{k49Corner}. The ellipsoids in the 2D marginal distribution are abruptly truncated and drop off to zero probability at $\mu_1 = 0$ or $\mu_2 = 0$, resulting in a non-Gaussian posterior. This is due to the prior, which only allows physically meaningful positive mass ratios. We again see strong correlations between planet 1 and 2 masses and eccentricity components. In Table \ref{k49Recover}, we can compare the median values recovered from the posterior to the true values. We can see that recovered values for masses of both planets are much more poorly constrained, with larger uncertainties than those of the Kepler-307 models, while the uncertainties for the eccentricity components are comparable. All of the true values fall within the central 68\% CI of the recovered values, but some have large uncertainties, only allowing them to be roughly constrained.

\subsubsection{Kepler-57 (KOI-1270) Dataset}

Lastly, we assess the performance of samplers on the TTVFaster model generated using true parameters close to those of the Kepler-57 system. Table \ref{DiagnosticTable} shows that  GAMC ($k=50,000$) and DEMCMC have the highest sampler efficiencies in terms of Min(ESS)/time. Again GAMC's efficiency appears to decrease after many iterations, but its value for Mean(ESS)/time is maximized, suggesting that without Hessian updates some parameters can still be sampled very efficiently, but others get stuck in certain regions of parameter space.
%move speculation to discussion?
 SMMALA and GAMC ($k=0$) perform poorly due to the long Hessian computation time, while AIMCMC and HMC ($n=7$) appear to perform comparatively better.

\begin{figure*}
\centering
\includegraphics[width=\textwidth]{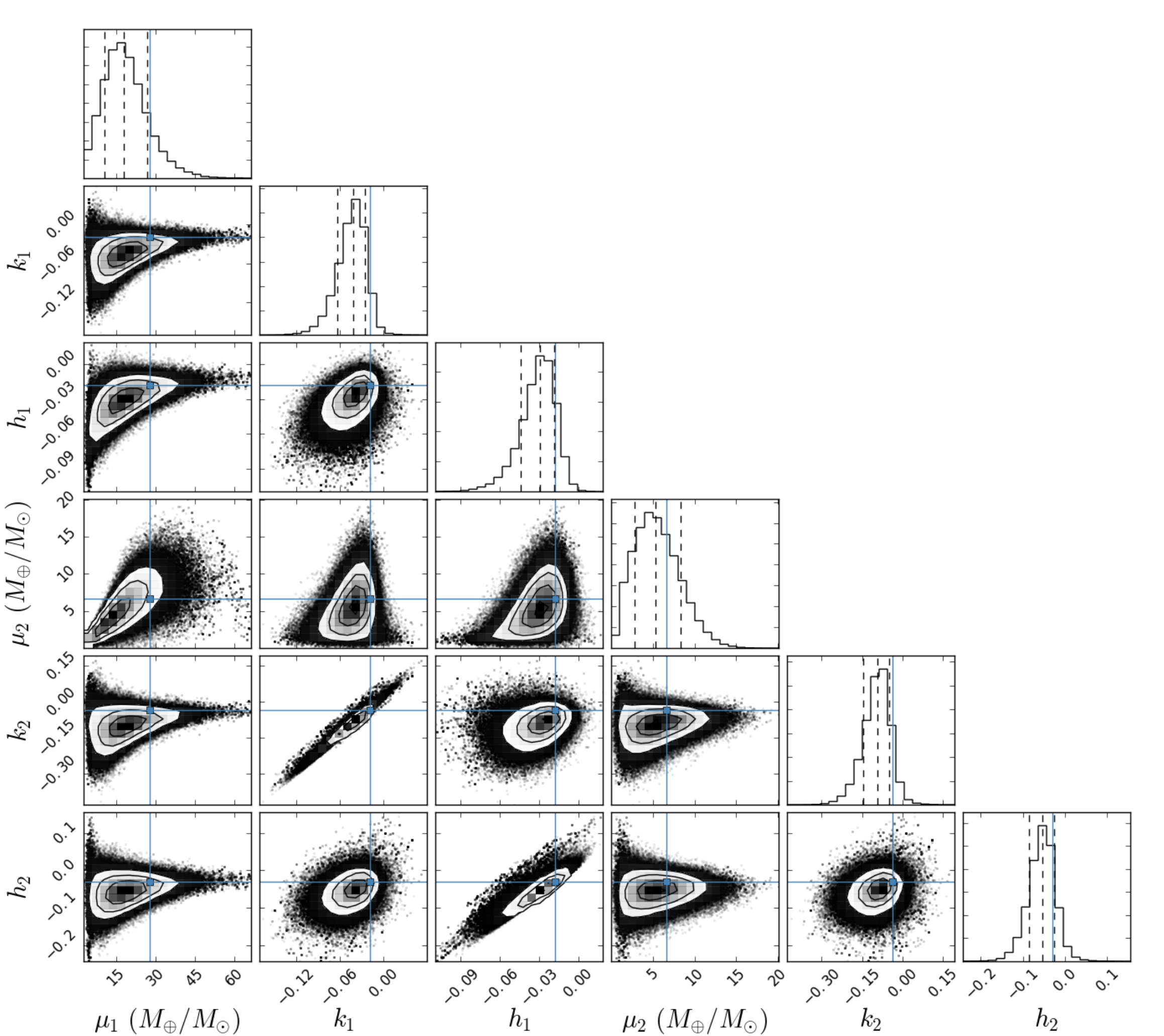}
\caption{Corner plot showing the posterior distribution for the Kepler-57 (KOI-1270) TTVFaster model. This chain used the DEMCMC sampler on a transformed parameter space with 30 walkers and ran for 150,000 generations with a 50,000 generation burn-in period.}
\label{k57Corner}
\end{figure*}

%modified for revisions
\begin{table*}
\caption{``True'' values used to generate the Kepler-57 dataset and summary statistics from the posterior distributions from both the DEMCMC and GAMC chains. Mass ratios and eccentricities reported here are unitless.}
\label{k57Recover}
\begin{tabular}{llll}
\hline
\textbf{Parameter} & \textbf{True Value} & \textbf{Recovered DEMCMC}& \textbf{Recovered GAMC}
\\ \hline
$\mu_1$               & $ 8.35 *10^{-5}$  &  $(5.4 ^{+2.7} _{-2.2})* 10^{-5}$ & 
$(5.4 ^{+2.6} _{-2.2})* 10^{-5}$
\\
$P_1$ (d)                  & 5.729500               & $5.729491 \pm 0.000008$ & 
$5.729490 \pm 0.000008$
\\
$t_{i,1}$ (d)             &  781.9966               & $781.9956 \pm 0.0005$ & 
$781.9956 \pm 0.0005$
\\
$k_1$                 &  -0.019                     &  $-0.042 ^{+0.016} _{-0.022}$ & 
$-0.042^{+0.016}  _{-0.021}$
\\
$h_1$                 &  -0.018                    & $-0.030 ^{+0.011} _{-0.015}$ & 
$-0.030 ^{+0.011} _{-0.015}$
\\
$\mu_2$               & $ 1.98799* 10^{-5}$  & $(1.6 ^{+0.9} _{-0.8})*10^{-5}$ &
$(1.6 ^{+0.9} _{-0.8}) * 10^{-5}$
\\
$P_2$ (d)                  &  11.60650                   & $11.60649 \pm 0.00002$ &
$11.60649 \pm 0.00002$
\\
$t_{i,2}$ (d)              &  786.7562                   & $786.7569\pm 0.0007$ &
$786.7569 \pm 0.0007$
\\
$k_2$                    &   -0.036                 & $-0.092 ^{+0.043} _{-0.053}$ &
$-0.091 ^{+0.042} _{-0.052}$
\\
$h_2$                    &  -0.030                    & $-0.054 ^{+0.028} _{-0.032}$ &
$-0.053 ^{+0.028} _{-0.032}$
\\ \hline
\end{tabular}

\end{table*}

We show the posterior sampled from running the DEMCMC sampler with 30 walkers for 150,000 generations in Figure \ref{k57Corner}. Many of the marginal distributions appear to be skewed, and the 2-D marginals do not appear to be simple Gaussian ellipsoids, but rather show irregular curvature and nonlinear dependences between parameters. Eccentricity vector components are still highly correlated, but masses have a weaker correlation than those of the other models. This bizarre posterior distribution caused us to question whether our MCMC chains had truly converged, so we decided to try running another long MCMC chain using GAMC, the second best performing sampler. The results for the posterior obtained from the GAMC chain, were near identical to those of DEMCMC, so it appears that both of these algorithms result in consistent posterior samples and their chains have come close to converging on the true posterior distribution. Table \ref{k57Recover} compares the true values for the model parameters to the values we recovered from both our DEMCMC and GAMC chains. Since the posteriors were almost exactly the same for both samplers, we only represent the DEMCMC chain on Figure \ref{k57Corner}.
%delete this sentence?
 For many of the parameters, such as the inner planet mass ratios and eccentricity $h$ and $k$ components, the true values don't fall within the 68\% CI of the recovered values, but rather appear just above its upper bound for most parameters. Since the masses and eccentricity components are highly correlated, it makes sense that a slight overestimate of one parameter would result in overestimates for the other parameters. The posterior for this dataset has a larger spread in values for the mass ratios than the other datasets, but eccentricities appear to have lower uncertainties and are better constrained than those of the other TTVFaster models.

\section{Discussion and Conclusions}
%k57 harder to recover values

%could try GAMC with other update schedule
%mu and h k from texture on TTV curve
In this study, we compared the efficiency of several MCMC methods used to sample from the posterior distributions of parameters for modeling exoplanet transit timing variations. We found that for all of our models, it helped tremendously to apply a linear transformation where we rotate and rescale the parameter space to obtain a transformed space in which the posterior is closer to a standard multivariate normal distribution (see Section \ref{sec:transform}). Without this transformation, the samplers which only use the target gradient, HMC and MALA, are only able to accept steps for tiny step sizes on the order of $10^{-6}$ or less. This renders these samplers ineffective, as they are unable to produce a large enough effective sample size in a reasonable number of iterations. The DEMCMC sampler actually performs relatively well in an untransformed space, but applying the transformation increases the ESS by an order of magnitude. The SMMALA and GAMC ($k=0$) samplers, which use Hessian updates most frequently, perform the best in the untransformed space in terms of Min(ESS)/time because knowledge of the target Hessian actually allows them to take reasonable step sizes. However, in the untransformed space, the GAMC sampler encounters numerical issues associated with Cholesky decompositions after many steps. When GAMC doesn't use use the target Hessian, it instead takes adaptive Metropolis steps. These steps use the Cholesky decomposition of the inverse of the last Hessian evaluation, but due to numerical roundoff errors, inverse Hessians are often not positive definite. Taking their Cholesky decompositions result in errors, and stabilizing against such errors risks distorting the transformation. Therefore, we use the coordinate transformation from Section \ref{sec:transform} to avoid these problems and compare all the samplers on equal footing.

Starting with a simple model for TTVs, comprised of a sinusoid with a first harmonic term added to a linear ephemeris for each planet (see Equation \eqref{SSMeq}), we assessed the performance of the different samplers in sampling from the posterior. This simple sinusoidal model (SSM) has a nearly Gaussian posterior with a few correlated parameters, and the HMC sampler with $n=2$ steps appears to have the best sampling efficiency. With such a posterior distribution, it is simple to apply the transformation and achieve a near Gaussian parameter space. Since the values of Min(ESS)/time in Table \ref{DiagnosticTable} are significantly larger than those of the other models, we conclude that this model is easier to sample from than the others.

  The TTVFaster models based on the Kepler-307 system had similar results. The posterior distributions were smooth and ellipsoidal enough that they can readily be rotated and scaled to appear nearly uncorrelated and close to a standard multivariate normal distribution. The model with 5 minutes of Gaussian white noise also had HMC ($n=2$) as the most efficient sampler, while the model with 15 minutes of Gaussian white noise found that MALA or HMC ($n=3$) performed best.

  For the Kepler-49 model, we found that the GAMC sampler performed the best after $k=50,000$ iterations.  This model is more difficult to sample from than the other models, presumably because the posterior distribution is truncated at zero masses, so no amount of  rotating and scaling can make the posterior appear Gaussian.

Lastly, for the Kepler-57 model, we found that both the DEMCMC sampler and GAMC after many iterations perform most efficiently in terms of Min(ESS) over wall clock time. The posterior distribution is skewed and several parameters exhibit nonlinear dependences with each other so it can't be transformed to appear Gaussian. Still, the best transformed parameter space for this model appears to be more easily sampled from than that of the Kepler-49 model, due to the larger step sizes that the chains can take.

For several of the models, it appeared that the performance of the GAMC sampler peaked after a certain number of iterations. We wanted to investigate this possibility, so we considered the GAMC sampler used to sample from the Kepler-57 dataset starting at a wider range of starting iteration values. Running 10,000 iteration pilot chains to diagnose efficiencies of chains starting at different points in the decay schedule, we didn't observe a clean curve where efficiency rises and drops off for different starting iterations. Instead we see that the efficiencies start off low at $k=0$  and gradually settle on their value at $k=10^6$. Between these two extremes, there are several seemingly random spikes with higher efficiency (such as $k=50,000$), but no clear smooth maximum for efficiency. This means that for longer MCMC chains, one should expect the efficiencies to be on average closer to the asymptotic values of efficiencies near $k=10^6$ rather than the high efficiency spikes earlier in the chain.

Overall, we have used a variety of MCMC methods to sample from TTVFaster models to accurately recover the true values for multiple synthetic datasets. The speeds at which the different samplers converge on the posterior distribution vary substantially, but our best performing samplers only take hours to converge as opposed to the weeks that would be required to obtain a large enough effective sample size for the worst performing samplers in the untransformed space.
No one sampler performs best for all datasets, but rather the relative efficiencies of samplers depend on how well the posterior can be transformed to appear Gaussian. In general, for a model such as TTVFaster, we would recommend applying our coordinate transform, detailed in Section \ref{sec:transform}, and using the GAMC or DEMCMC samplers. GAMC doesn't always perform better than the other samplers, but after more that 25,000 iterations, it consistently obtains large enough effective sample sizes to converge relatively rapidly. Similarly the DEMCMC sampler performs well, though not necessarily at the top, for all of the datasets, and its insensitivity to tuning parameters bypasses the difficulties found in tuning the other samplers. For all of our datasets, we observe that DEMCMC performs substantially better than AIMCMC, and we would recommend the use of the DEMCMC sampler above AIMCMC for the purpose of the characterization of exoplanet TTVs.

In future studies, we would like to investigate how the length of the burn-in period for MCMC chains is affected by our choice of initial parameter values and which sampler we use. It is often the case that one doesn't have concrete initial guesses for model parameters other than the periods and times of first transits, so we would like to determine how rapidly each algorithm can find the region near the true values, with a sub-optimal starting position. We also aim to apply our MCMC methods to actual TTV datasets, and we  would eventually like to assess the performance of these samplers for more precise and computationally intensive N-body models. With more efficient MCMC sampling methods, we hope to  greatly reduce the computation time needed to characterize planetary systems using TTV models.

\section*{Acknowledgements}

EBF and NWT acknowledge the support of the Institute for CyberScience and the Center for Exoplanets and Habitable Worlds, which is supported by The Pennsylvania State University, the Eberly College of Science, and the Pennsylvania Space Grant Consortium. EBF and NWT were supported from NSF grant AST1616086, and NWT acknowledges support from the Penn State Center for Astrostatistics.

\bibliographystyle{mnras}
\bibliography{ttvsources}

\appendix
\section{GAMC Sampler}
\label{appendix:GAMC}

The geometric adaptive Monte Carlo (GAMC) sampler was recently introduced by \citet*{papamarkou2018}.
This appendix provides a brief overview of GAMC.

The main goal of GAMC is to balance the exploitation of the local geometry of the parameter space with total computational
time. This act of balance is achieved by sampling in a random environment.
The random environment regulates the switching between local geometric and adaptive
proposal kernels via a schedule.
An exponential schedule enables more frequent use of local geometric information in early iterations of the chain,
while saving computational time in late iterations.
The average complexity can be manually set via a hyperparameter associated with the exponential schedule.

GAMC relies on a sequence $\{B_k\}$ of IID binary random variables, called the random environment.
$B_k$ follows Bernoulli distribution with probability
$s_k := P(B_k = 1)$.
Depending on whether $B_k=1$ or $B_k=0$, the proposal kernel at the $k$-th iteration of GAMC is set to a
geometric Langevin Monte Carlo or adaptive Metropolis proposal kernel, respectively.
The sequence of probabilites $\{s_k\}$ determines the frequency of using a geometric kernel.
In this paper, the probabilities $s_k$ is set to
\begin{equation}
\label{exp_schedule}
s_k = \exp{(- rk)},
\end{equation}
where $r$ is a positive-valued tuning hyperparameter.
Larger values of $r$ in \eqref{exp_schedule} result in faster reduction in the probability
of taking geometric steps.
For the GAMC samplers we test, we hold $r$ constant at $r=2.0 * 10^{-5}$ and observe their performance starting at the $k$-th iteration.

\bsp	% typesetting comment
\label{lastpage}
\end{document}